\documentstyle[twoside,fleqn,espcrc2,rotate,epsf]{article}
\newcommand{\be}{\begin{equation}}
\newcommand{\ee}{\end{equation}}
\newcommand{\bear}{\begin{eqnarray}}
\newcommand{\ear}{\end{eqnarray}}
\newcommand{\slp}{\raise.15ex\hbox{$/$}\kern-.57em\hbox{$p$}}
\newcommand{\sleta}{\raise.15ex\hbox{$/$}\kern-.57em\hbox{$\eta$}}
\newcommand{\slK}{\raise.15ex\hbox{$/$}\kern-.57em\hbox{$K$}}
\newcommand{\slv}{\raise.15ex\hbox{$/$}\kern-.57em\hbox{$v$}}

\title{Form Factors for Semi-Leptonic
and Radiative Decays of Heavy Mesons to  Light Mesons}

\author{Berthold Stech\\
Institut f\"ur Theoretische Physik, Universit\"at Heidelberg\\
Philosophenweg 16, D-69120 Heidelberg\\
E-mail: B.Stech@ThPhys.Uni-Heidelberg.DE}

\begin{document}

\begin{abstract}
To know and understand form factors of hadronic currents is of
decisive importance for analysing exclusive weak decays. The
ratios of different form factors of a given process depend
on the relativistic spin structure of initial and final particles.
It is shown --- assuming simple properties of the spectator
particle --- that these ratios can entirely be expresssed in
terms of particle and quark mass parameters.
For quark masses large compared to the spectator mass the
Isgur-Wise relations follow. The corresponding amplitudes for
heavy-to-light transitions show a very similar structure. In
particular, the $F_0$ and $A_1$ form factors behave again differently
from the $F_1, A_2, V$ and $T_1$ form factors.
\end{abstract}

\maketitle
%\twocolumn
 The subject of this talk concerns exclusive semileptonic and radiative
decays of heavy mesons. These decays play an outstanding role for
the determination of the parameters of the standard model, in particular
the quark-mixing parameters. They are also an important source of
information about the still badly understood QCD dynamics in the
confinement region. The dynamical content of the corresponding amplitudes
is contained in Lorentz-invariant form factors. The
calculation of these form factors requires a
non-perturbative treatment. Many theoretical tools have been applied
for their determination: quark models, QCD sum rules, and lattice
calculations. The machinery of sum rules and lattice calculations
has the advantage to be directly based on the QCD Lagrangian, but is
quite involved. Quark models, on the other hand, are less directly
connected with the QCD Lagrangian, but give
a vivid picture of what is going on and allow an easy application
to different processes and quite different kinematical regions.
However, models presented so far lacked full relativistic
covariance with respect to quark spins, and it is the spin structure
which determines the ratio of different form factors we are concerned
with here. The knowledge of these ratios is of paramount importance
for the analysis of experimentally measured decays.

The problem of the ratios of different form factors is solved
in the formal limit where initial and final
quark masses are taken to be infinite. For instance, for $m_c\to
\infty, m_b\to\infty$ the two form factors $(F_1,F_0)$ describing
$B\to D$ transitions and the 4 form factors $(V,A_0,A_1,A_2)$
describing $B\to D^*$ transitions are all related \cite{1},\cite{2}:
\bear\label{1}
&&F_1=V=A_0=A_2\nonumber\\
&&A_1=F_0=(1-q^2/(m_B+m_D)^2)F_1.\ear
Denoting initial and final meson masses by $m_I$ and
$m_F$, respectively, the Isgur-Wise function $\xi(y)$
connected to $F_1$ by
\be\label{2}
F_1=\frac{1}{2}\sqrt{\frac{m_I}{m_F}}
\left(1+\frac{m_F}{m_I}\right)\xi(y)_{Isgur-Wise}\ee
depends in this limit on the product of the
4-velocities $v_I$ and $v_F$ only
\be\label{3}
y=v_F\cdot v_I=\frac{m_I^2+m_F^2-q^2}{2m_Im_F}.\ee

The prediction of the corrections
to the simple relations (1) for the case of finite physical quark
masses
are a challenge to quark models.
The calculation of form factors is even
more challenging for heavy-to-light transitions such
as
\be\label{4}
\bar B^0\to\pi^+\ e^-\bar\nu_e,\quad\bar B^0\to\rho^+
\ e^-\bar\nu_e.\ee
The flavor and spin symmetries necessary to derive (1) do not apply in
this case. Only the scaling property of the matrix elements with
respect to
the mass $m_I$ can be used. It connects, for example, $B\to F$
with $D\to F$ transitions. But this is of little help since the
application is restricted to a limited kinematic region
near $q^2=q^2_{max}$ and holds strictly only in the limit
of large masses.

In the following I will use a fully relativistic quark model
approach to get form factor relations.
These relations follow without a detailed knowledge about quark
model wave functions and should, therefore, be valid in the
framework of a large class of dynamical models.
They depend on mass parameters
only and will reduce to eq. (1) in the limit of large masses
of the active quarks.

Let us consider the 4-momentum of an initial meson $(I)$ with
velocity $v_I$ and divide it into the momenta of the active
quark $(i)$ in this meson and the spectator $(sp)$
\bear\label{5}
&&P^I=p^I_i+p^I_{sp}\nonumber\\
&&p^I_i=\epsilon^I_iv_I+k_I,
\quad p_{sp}^I=\epsilon^I_{sp}v_I-k_I\nonumber\\
&&\epsilon^I_i+\epsilon^I_{sp}=m_I.\ear
The dynamics of the bound state will be described by a
momentum space wave function $f_I(k_I,v_I)$. Its peak is taken
to be at $k_I=0$ defining, thereby, the splitting of the meson
mass into the two constituent masses. Choosing the simplest internal
$S$-wave structure, the wave function for the decaying
pseudoscalar meson has
the form (in coordinate space)
\bear\label{6}
&&\psi_I(x_1,x_2)=\frac{1}{N_I^{1/2}}\int d^4k_If_I(k^2_I,v_I
\cdot k_I)\times\nonumber\\
&&(\slp^I_i+m_i)\gamma_5(m_{sp}-\slp^I_{sp})e^{-ip^I_ix_1}
e^{-ip^I_{sp}x_2}.\ear
$\psi_I$ is a $4\times4$ matrix. The mass values $m_i, m_{sp}$
appearing in the propagators
are not necessarily identical with the constituent masses
$\epsilon_i^I,\epsilon^I_{sp}$. Expressions analog to (5,6) hold
for the final particle (denoted by $F$) emitted in the decay
process. If it is a vector particle --- a $\rho$ meson for instance
--- the Dirac matrix $\gamma_5$ has to be replaced by the polarisation
matrix $\sleta$ of this particle.

The $I\to F$ transition amplitude is obtained from the product $\psi
_I\bar\psi_F$. The integration over the space-time coordinate
$x_2$ of the spectator particle
leads to momentum conservation of this particle $p^F_{sp}=p^I
_{sp}$ correlating thereby the momentum $k_F$ with $k_I$:
\be\label{7}
k_F-\epsilon^F_{sp}v_F=k_I-\epsilon_{sp}^Iv_I\equiv K.\ee
The off-shell quark momenta can thus be written
\bear\label{8}
&&p_i^I(K)=m_Iv_I+K,\quad p^I_{sp}=p_{sp}^F=-K,\nonumber\\
&&p^F_f(K)=m_Fv_F+K.\ear
The wave functions $\psi_I,\psi_F$ contain the propagators
of the active quarks and of the spectator. Because the weak current
does not act on the spectator, the transition amplitude should
contain this propagator only once as is evident from the
corresponding triangle graph \cite{3}. Thus, by inserting
the inverse propagator of the spectator, the transition
amplitude reads
\bear\label{9}
&&<F(v_F)|(\bar q_f(0)\Gamma q_i(0))|I(v_I)>=\nonumber\\
&&\frac{(2\pi)^4}{(N_FN_I)^{1/2}}\int d^4Kf^*_F((K+\epsilon^F_{sp}
v_F)^2,\nonumber\\
&&v_F\cdot K+\epsilon_{sp}^F)\cdot(m^2_{sp}
-p^2_{sp})\cdot
\nonumber\\
&&f_I((K+\epsilon_{sp}^I\cdot v_I)^2,\ v_I\cdot K+\epsilon^I_{sp}))
\cdot J(K)\ear
with
\bear\label{10}
&&J(K)=Spur\left\{\Gamma(\slp^I_i(K)+m^I_i)\gamma_5\right.\nonumber\\
&&\left.(m_{sp}+\slK){\gamma_5 \choose \sleta^*}
(\slp_f^F+m_f^F)\right\}.
\ear
In semileptonic decays $\Gamma$ stands for $\gamma_\mu(1-\gamma_5)$; in
the calculation of $I\to F\gamma$ transitions it
stands for $\sigma_{\mu\nu}q^\nu(1+\gamma_5)$ (apart from a global factor).

The initial and final wave functions have their maximum at $k_F=
k_I=0$ with a width corresponding to the particle sizes.
The integrand of the transition amplitude, on the other hand,
has its maximum at values of $k_B$ and $k_F$ different from zero.
One can expect this maximum to occur at $k_I=\bar k_I,
k_F=\bar k_F$ with
\be\label{11}
v_I\cdot\bar k_I=\frac{1}{2}(\epsilon^I_{sp}-\epsilon^F_{sp}
)=-v_F\cdot\bar k_F.\ee
The reason is that, together with (7,8), (11) implies
small and equal off-shell
values for all three propagators occuring in the transition
matrix element, as well as average quark energies close to their
constituent masses in the rest system of the relevant particles:
Because the average transverse components of
$p_{sp}=-K$ vanish, one gets with
$\epsilon_{sp}=(\epsilon^I_{sp}+\epsilon^F_{sp})/2$
from (7) and (11)
\be\label{12}
\bar p_{sp}=-\bar K=\epsilon_{sp}\frac{v_I+v_F}{1+y}\ee
and thus
\bear\label{13}
&&\bar p^I_i\cdot v_I=m_I-\epsilon_{sp}\nonumber\\
&&\bar p_{sp}\cdot v_I=\bar p_{sp}\cdot v_F=\epsilon_{sp}\nonumber\\
&&\bar p^F_f\cdot v_F=m_F-\epsilon_{sp}\nonumber\\
&&(\bar p_i^I)^2-(m_I-\epsilon_{sp})^2=(\bar
p_{sp})^2-\epsilon_{sp}^2=\nonumber\\
&&(\bar p^F_f)^2-(m_F-\epsilon_{sp})^2=-\epsilon_{sp}^2\frac{y-1}{y+1},
\nonumber\\
&&\bar k^2_I=\bar k^2_F=-\epsilon_{sp}^2\frac{y-1}{y+1}+
\biggl(\frac{\epsilon^I_{sp}-\epsilon^F_{sp}}{2}\biggr)^2.\ear
According to (12) the average space velocity of the spectator
vanishes in the special coordinate system where $\vec v_F
=-\vec v_I$ as one would expect. In the following I will assume
(12) to hold and to decisively determine the structure of the
transition amplitude.

Considering the quark momenta (eq. (8))
in the physical region of the variable $y=v_F\cdot v_I$
and taking $\bar K$ for $K$, it is seen that
at the maximum of the transition amplitude the decaying
and emitted quarks carry essentially the same momenta as the mesons
they are part of. This is especially true in heavy-to-heavy
transitions where $m_I,m_F\gg\epsilon_{sp}$, but holds also
in heavy-to-light decay processes at least for
large values of $y$ (i.e. low $q^2$-values) \cite{4}.

The covariant structure of the transition amplitude is obtained
from the integral over $J(K)$ in (9). For wave functions
with a strong peak in momentum space and a width of
order of the constituent mass of a light quark one may replace
$J(K)$ by $J(\bar K)$. This replacement saves us from an
unfruitful discussion of specific wave
functions or propagators in the confinement region which
cannot be reliably calculated at present. But it is certainly
an approximation which holds good only for strongly peaked
and otherwise smooth wave functions.

It is now a straightforward task to decompose
\bear\label{14}
&&J(\bar K)=Spur\{\Gamma(m_I\slv_I+\bar {\slK}+m^I_i)\gamma_5\cdot
\nonumber\\
&&(m_{sp}+\bar {\slK}){\gamma_5\choose \sleta^*}(m_F\slv_F+\bar{\slK}+m^F_f)\}
\ear
in terms of covariant expressions and to extract the
corresponding form factors. There remains, of course, an undetermined
function of the variable $y$ multiplying the form factors.
The result can be written in the form
\bear\label{15}
&&F_1=\rho^{FI}(y)(1+\zeta^{FI}_{F1}(y))\nonumber\\
&&F_0=\rho^{FI}(y)(1-\frac{q^2}{(m_I+m_F)^2})(1+\zeta^{FI}_{F_0}(y))
\nonumber\\
&&\nonumber\\
&&V=\rho^{FI}(y)(1+\zeta^{FI}_V(y))\nonumber\\
&&A_1=\rho^{FI}(y)(1-\frac{q^2}{(m_I+m_F)^2})(1+
\zeta_{A_1}^{FI}(y))\nonumber\\
&&\nonumber\\
&&A_2=\rho^{FI}(y)(1+\zeta^{FI}_{A_2}(y))\nonumber\\
&&A_0=\rho^{FI}(y)(1+\zeta^{FI}_{A_0}(y))\nonumber\\
&&T_1=\rho^{FI}(y)(1+\zeta_{T_1}^{FI}(y)).\ear
Here $T_1$ is defined as the form factor relevant for
radiative decays:
\bear\label{16}
&&<F^*|(\bar q_f\sigma_{\mu\nu}
(1+\gamma_5)q^\nu q_i)|I>=\nonumber\\
&&\epsilon_{\mu\nu
\lambda\rho}\eta^{*\nu}_FP^\lambda_I P^\rho_F2T_1-\nonumber\\
&&i\,(\eta^*_\mu(m^2_I-m_F^2)-(\eta^*\cdot P^I)(P^I+P^F)_\mu)T_2-\nonumber\\
&&i\,(\eta^*\cdot P^I)\cdot\nonumber\\
&&\left((P^I-P^F)_\mu-\frac{q^2}{m_I^2-m^2_F}
(P^I+P^F)_\mu\right) T_3,\nonumber\\
&&T_2(q^2=0)=T_1(q^2=0)\,,\quad \epsilon_{0123}=1 \ear
The functions $\zeta^{FI}(y)$ depend on dimensionless
combinations of the masses contained in (14). They all
vanish in the limit
\be\label{17}m_i,m_I,m_f,m_F\gg m_{sp},\epsilon_{sp}.\ee
Thus, (\ref{15}) contains the Isgur-Wise result eq. (1).

For the general case in which (\ref{17}) does not hold,
the functions $\zeta^{FI}(y)$ could be written down
but are too long to be displayed here.

The constituent quark picture suggests to use in these
expressions (for a light quark spectator)
\be\label{18}
m_i=m_I-\epsilon_{sp},\ m_f=m_F-\epsilon_{sp},\ m_{sp}=\epsilon_{sp}
\ee
Then, the functions $\zeta^{FI}(y)$ depend, besides upon $y$
and the particle mass ratio $m_F/m_I$, upon the value of
$\delta_F=\epsilon_{sp}/m_F$ only.\footnote{The use of current
mass values, for instance $m_b\simeq4.8$ GeV, $m_u\approx
0$ is also conceivable. Choosing in addition $m_{sp}\approx0$, eq. (11)
of ref. \cite{4} can be rederived.}
Since in $B\to D^*,B\to\rho$ and $D\to K^*,D\to\rho$ transitions
$\epsilon^I_{sp}
\approx\epsilon_{sp}^F\approx 0.35$ GeV appears to be a reasonable
value, all $\zeta^{FI}$ functions are predictable using the
corresponding values for $\delta_F$.
As an example the functions $1+\zeta^{\rho B}(y)$ for the
semi-leptonic $B\to\rho$ transitions and the function
$1+\zeta^{\rho B}_{T_1}$ are plotted in Fig. 1. ($\zeta_{T_1}$
turned out to be identical to $\zeta_{F_1}$ independent of the
assumption (18)).
\begin{figure}[thb]
\centerline{
\rotate[r]{
\epsfysize=3in
\epsffile{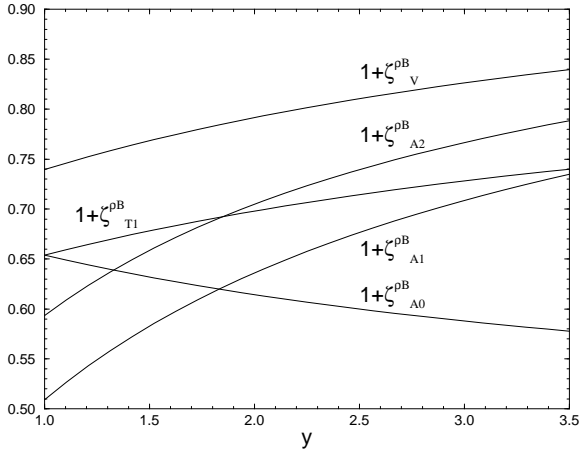} }}
\caption[]{The functions $1+\zeta^{\rho B}$ for semileptonic and radiative
$B\rightarrow\rho$ decays. Quark masses according to Eq. (18) and
$\epsilon_{sp}=0.35\,{\rm GeV}$.}
\end{figure}
Because the curves for $\zeta_V,\zeta_{A_1}$ and
$\zeta_{A_2}$ run similarly it is seen  from (15) that the
form factor $A_1$ differs in its $q^2$ dependence from the $V$ and
$A_2$ form factors, like it is the case in the heavy quark limit.
It is a consequence of relativistic covariance \cite{5}, \cite{4}.

In Fig. 2 the longitudinal polarization of the $\rho$-meson is shown
using again (18) and $\epsilon_{sp}=0.35$ GeV.
For comparison, the polarization without mass correction (i.e. with
$\zeta^{\rho B}=0)$ is also
plotted.
\begin{figure}[thb]
\centerline{
\rotate[r]{
\epsfysize=3in
\epsffile{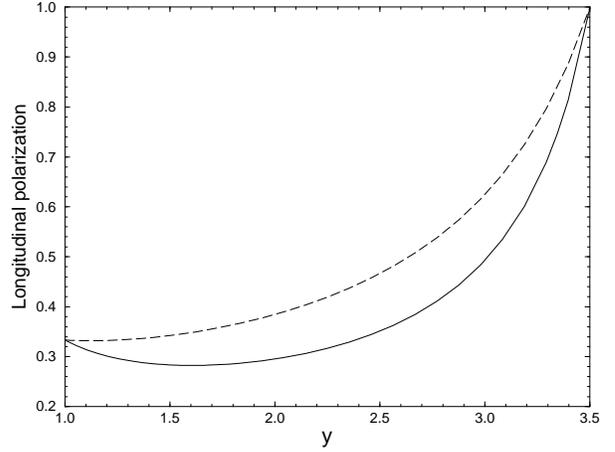} }}
\caption[]{Longitudinal polarization of the $\rho$-meson in semileptonic
$B\rightarrow\rho$ transitions. Full line: quark masses according to Eq. (18)
and $\epsilon_{sp}=0.35\,{\rm GeV}$. Dashed line: without mass corrections
($\zeta^{\rho B}=0$).}
\end{figure}

Because of the lack of spin symmetry in the light sector a relation
between, for instance, the $B\to\rho$ and $B\to\pi$ form factors
cannot be obtained from (15) even though the functions $\zeta^{FI}$
are known. The factor $\rho^{FI}(y)$ depends on the particle masses and
on the internal structures of the initial and final particles. The
explicit dependence on the particle masses can be taken care of, however,
by setting
\be\label{19}
\rho^{FI}=\frac{1}{2}\sqrt{\frac{m_I}{m_F}}(1+\frac{m_F}{m_I})
\xi^{FI}(y).\ee
The form factors as obtained from (15) and (19) have now the correct
scaling property for fixed $y$. Furthermore, for large masses of the active
quarks $\xi^{FI}(y)$ defined by (19) turns into the normalized
Isgur-Wise function
\be\label{20}
\xi^{FI}(y)\to \xi_{Isgur-Wise}(y).\ee
For phenomenological applications we can specify $\xi^{FI}(y)$ further:
\be\label{21}
\xi^{FI}(y)=\sqrt{\frac{2}{y+1}}\left(\frac{1}{2}+
\frac{1}{1+y}\right)g^{FI}\Big(\bar k^2_I(y)\Big),\ee
\begin{displaymath}
\bar k^2_I(y)=\bar k^2_F(y)=-\epsilon_{sp}^2\frac{y-1}{y+1}+
\biggl(\frac{\epsilon^I_{sp}-\epsilon^F_{sp}}{2}\biggr)^2.\end{displaymath}
The first factor in (21) is necessary to give mass independence
of the form factors in the limit $m_I/
m_F\to\infty$ at fixed $q^2$. The second factor is obtained from $J(\bar K)$
by setting $m_{sp}=\epsilon_{sp}$. It was divided out in defining
the Isgur-Wise limit. The function $g^{FI}$ depends on the variable
$\bar k^2_I=\bar k^2_F$ obtained in (\ref{13}) and is a increasing
function of this variable.

As an illustration of the usefulness of (21) one may take a simple
dipole formula which contains then just one parameter:
\be\label{22}
g^{FI}=\Big(1-x^{FI}\bar k^2_I(y)/m_{sp}^2 \Big)^{-2}.\ee
$g^{FI}$, or in the case of eq. (22), the parameter $x^{FI}$ depends
on the internal structure of initial and final states. Only in a
hypothetical world where the internal meson wave functions
are identical, $g^{FI}$ would be process-independent
and $\xi^{FI}(y)=\xi(y)$ a truly universal function describing
a large number of form factors. The difference from
this hypothetical world
seems, however, not to be a drastic one: Taking $x=0.5$,
the eqs. (21,22)
provide for a  reasonable Isgur-Wise function for
$B\to D^{(*)}$ decays. Together with (15), (19) one
gets for the branching ratio $BR(B\to D^*e^-\bar\nu_e)\approx7\%$
consistent with the experimental value \cite{6}. The same equations
also lead to $T_1(q^2=0)\approx0.36$ for the $B\to K^*\gamma$ process,
a value also obtained in QCD sum rule estimates \cite{7}. Applied
to the semileptonic $D\to K^*$ decay I obtain the branching ratio
$BR(D^0\to\bar K^*e^+\nu)\approx2\%$ also in accord with the data
\cite{6}. For the semileptonic $\bar B^0\to\rho^+e^-\bar\nu_e$
transition the form factors $V^{\rho B}, A_1^{\rho B}, A_2^{\rho B},
A_0^{\rho B}$ at $q^2=0$ turn out to be 0.37, 0.32, 0.35, 0.26,
respectively, in accord with QCD sum rule results \cite{7} and earlier
estimates \cite{8}.
The differential branching ratio for the $B\to\rho
e^-\bar\nu$ transition is plotted in Fig. 3 - taking for the $B$-meson
lifetime 1.5 psec and dividing by $|V_{ub}|^2$. Integrating
it one finds the branching ratio 21.1$|V_{ub}|^2$. (The
transitions to the light pseudoscalars $\pi$ and $K$ require a
more detailed treatment because the pole position is of
greater importance; (18) is not applicable and $\epsilon^I_{sp}
\not= \epsilon^F_{sp}$.)
\begin{figure}[thb]
\centerline{
\rotate[r]{
\epsfysize=3in
\epsffile{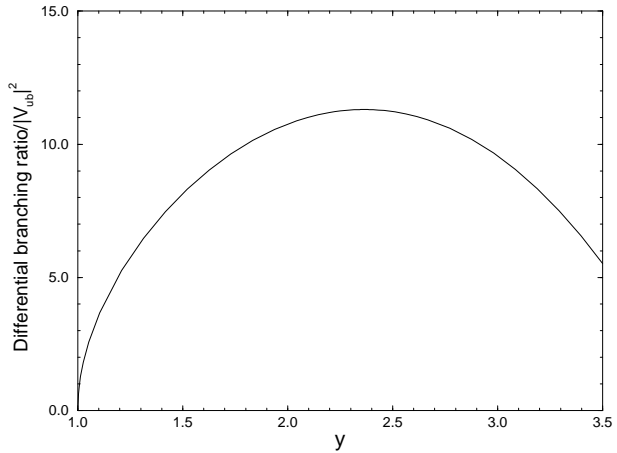} }}
\caption[]{Differential branching ratio/$|V_{ub}|^2$ for the
semileptonic $B\rightarrow\rho$ transition using $\xi(y)$ as described in
the text.}
\end{figure}

It should be clear that the ``results'' obtained from the Ansatz (22)
and by taking an unjustified universal value for the parameter
$x^{FI}$ serve as an illustration only and are not based on a detailed
analysis.

\bigskip
The author takes pleasure in thanking Dieter Gromes and Matthias Jamin
for useful discussions. He also likes to thank A. Fridman and the organizers
of the Strasbourg Symposium for the very pleasant meeting.


\begin{thebibliography}{99}
\bibitem{1} N. Isgur and M. B. Wise, Phys. Lett. {\bf B232}
(1989) 113; {\bf B237} (1990) 527.
\bibitem{2} M. Neubert and V. Rieckert, Nucl. Phys. {\bf B 382}
(1992) 97.
\bibitem{3} S. Mandelstam, Proc. Roy. Soc. {\bf 233} (1955) 248.
\bibitem{4} B. Stech, Phys. Lett. {\bf B354} (1995) 447.
\bibitem{5} R. Aleksan, A. Le Yaouanc, L. Oliver, P. P\`ene, and
J.-C. Raynal, LPTHE-Orsay 94/15, hep-ph 9408215.
\bibitem{6} Particle Data Group, Review of Particle Properties,
Phys. Rev. {\bf D50} (1994) 1173.
\bibitem{7} For references see V. M. Braun, Nordita preprint
hep-ph 9510404, Oct. 95.
\bibitem{8} M. Bauer, B. Stech, M. Wirbel, Z. Phys. {\bf C34}
(1987) 103.
\end{thebibliography}
\end{document}